# Molecular orbital formation and metastable short-range ordered structure in VO₂

Shunsuke Kitou[1,*], Akitoshi Nakano[2], Masato Imaizumi[2], Yuiga Nakamura[3],
Ichiro Terasaki[2], Taka-hisa Arima[1,4]

[1]*Department of Advanced Materials Science, The University of Tokyo, Kashiwa 277-8561, Japan.*
[2]*Department of Physics, Nagoya University, Nagoya 464-8602, Japan.*
[3]*Japan Synchrotron Radiation Research Institute (JASRI), SPring-8; Hyogo 679-5198, Japan.*
[4]*RIKEN Center for Emergent Matter Science, Wako 351-0198, Japan.*

The metal-insulator transition (MIT) in vanadium dioxide VO₂ due to V-V dimerization has been extensively discussed for decades. While it is widely acknowledged that electron correlations, Peierls instabilities, and molecular orbital formations are crucial for understanding the MIT of VO₂, the primary origin of the MIT remains controversial. In this study, we delve into the crystal structure and orbital state of VO₂ through synchrotron x-ray diffraction experiments. The molecular orbital formation corresponding to the V-V dimerization is directly observed in the low-temperature insulating monoclinic phase, called the M1 phase, as indicated by the valence electron density distribution. Moreover, diffuse scattering observed in the high-temperature metal phase of rutile structure suggests the presence of short-range correlation of V displacements, which is not directly attributed to the structural fluctuation of the M1 phase. The short-range order in the rutile phase will be the key to understanding the MIT in this system.

*Introduction.* Transition metal compounds with the orbital degree of freedom are an attractive platform for realizing molecular clusters [1]. The crystal structure and physical properties change dramatically with the formation of molecular clusters such as trimers in LiVO₂ [2], tetramers in K₂Mo₈O₁₆ [3], and octamers in CuIr₂S₄ [4]. Vanadium dioxide (VO₂) exhibiting a first-order metal-insulator transition (MIT) near room-temperature [5-7] is one of the most famous materials that form such molecular clusters [8-11]. The rutile-type tetragonal structure with the space group $P4_2/mnm$ in the metallic phase [12,13] changes to the monoclinic structure with the space group $P2_1/c$ in the insulator phase (M1 phase) [8-11] at $T_{MI} \cong 340$ K with decreasing temperature. In the rutile phase, the V atoms are arranged at equal intervals on straight lines along the $c_r$-axis [Fig. 1(a)], where a V⁴⁺ ion is located at the octahedral site surrounded by six O²⁻ ions. In the M1 phase, V atoms form zigzag V-V dimers [Fig. 1(b)] to realize a non-magnetic singlet state [6,7]. Although numerous studies discuss the importance of electron correlations [14-16], electron-lattice (Peierls) instabilities [17-19], and molecular orbital formations [20-23] in this system, the primary origin of the MIT is still controversial.

The most direct way to address the long-standing problem is to gain a proper understanding of the crystal structure and orbital state of VO₂. X-ray diffuse scattering (XDS) was observed around the R and M points in the first Brillouin zone in the rutile phase [24-26], indicating the presence of short-range correlations of atomic displacement. Previous extended x-ray absorption fine-structure spectroscopy found the short and long V-V bonds in the rutile phase [27]. However, it is not clear what kind of short-range ordered structure is realized in the rutile phase and how it relates to the MIT in VO₂. Moreover, the short-range correlation in the rutile phase may be related not only to another insulator phase induced by small impurity doping, called the M2 phase [28,29], but also to the transient disordering state induced by photoexcitation [30-40].

In this study, we perform synchrotron x-ray diffraction (XRD) experiments using single crystals of VO₂. The valence electron density (VED) analysis observes the molecular orbital formation of V $3d$ electrons in the M1 phase. We also observe two-dimensional plane-shaped XDS corresponding to the V displacement in the rutile phase, which is not ascribable to the structural fluctuations derived from the M1 phase.

*Experiments.* Single crystals of VO₂ were obtained by reducing molten V₂O₅ powder at 950°C in a flow of N₂ gas [41]. The MIT temperatures, $T_{MI} = 340$ and 336 K during the heating and cooling processes, were confirmed through resistivity measurements (Fig. S1 [42]), which are consistent with the previous reports [5-7]. XRD experiments were performed on BL02B1 at the SPring-8 synchrotron facility in Japan [43]. Small and large single crystals of $45 \times 40 \times 30$ and $140 \times 120 \times 120$ μm³ were used for the electron density analysis and the XDS observation, respectively. No monoclinic twins were present in the smaller single crystal. An N₂-gas-blowing device was employed for the measurements from 400 to 100 K, with a potential temperature error of approximately 5 K. The x-ray wavelength $\lambda$ was 0.30946 Å. A two-dimensional detector CdTe PILATUS was used to record the diffraction pattern. The intensities of Bragg reflections with the interplane distance $d > 0.28$ Å were collected by the CrysAlisPro program [44] using a fine slice method, in which the data were obtained by dividing the reciprocal lattice space in an increment of $\Delta\omega = 0.01°$. Intensities of equivalent reflections were averaged and the structural parameters were refined by using Jana2006 [45]. To extract the VED distribution around each atomic site, a core differential Fourier synthesis (CDFS) method was used [46,47], which has been applied to various strongly correlated electron materials such as titanate [47], vanadate [48], ferrate [49], and molybdate [50]. [Ar]- and [He]-type electron configurations were regarded as core electrons for V and O atoms, respectively. V $3d$ and O $2s/2p$ valence electrons should remain after the subtraction of the core electron density distribution. Crystal structure and VED distribution are visualized by using VESTA [51].

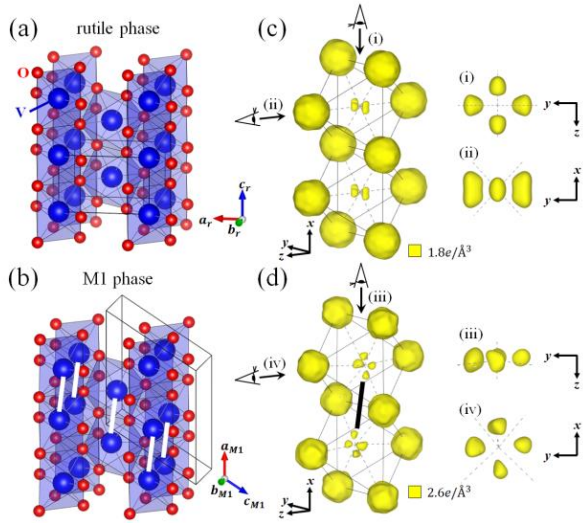

FIG. 1. Crystal structure of $VO_2$ at (a) 400 K in the rutile phase and (b) 100 K in the M1 phase. VED distributions obtained from the CDFS analysis at (c) 400 and (d) 100 K. Yellow iso-density surfaces show electron-density levels of 1.8 and $2.6e/\text{Å}^3$ at 400 and 100 K, respectively. Panels (i)-(iv) are magnified top and side views of the VED around the V site. White and black lines in (b) and (d) correspond to V-V dimers, respectively. $x$, $y$, and $z$ indicate the quantization axes, which are defined as $x \parallel c_r$, $y \parallel a_r - b_r$, and $z \parallel a_r + b_r$. $a_r$, $b_r$, and $c_r$ indicate the unit cell axes of the rutile phase.

Figures 1(c) and 1(d) show the VED distributions at 400 and 100 K, respectively. The valence electrons are observed around the V and O sites, as shown by yellow iso-density surfaces. The isotropic VED distributions around the O site are consistent with the $2s^2 2p^6$ electron configuration. On the other hand, anisotropic VED distributions are clearly observed around the V site, which corresponds to the $3d^1$ electron of a $V^{4+}$ ion. The anisotropy is different between the rutile and M1 phases, suggesting a change of the $3d$ orbital state with dimerization.

*Results and Discussion.* Figures 1(a) and 1(b) show the crystal structure at 400 K (rutile phase) and 100 K (M1 phase), respectively. Here, the unit-cell axes of each phase have approximate relationships of $a_{M1} \parallel c_r$, $b_{M1} \parallel b_r$, and $c_{M1} \parallel -a_r - c_r$, where the subscripts $r$ and $M1$ denote the rutile and M1 phases, respectively. The V-V bond length along the $c_r$-axis is 2.8565(3) Å in the rutile phase, whereas the V-V short and long bond lengths in the M1 phase are 2.60546(13) and 3.16875(14) Å, respectively. The short bond corresponds to the V-V dimer, which is tilted by 6.9450(15)° from the $a_{M1}$-axis. These results are consistent with previous reports [8–13]. Details of the structural parameters are summarized in Tables S1-S4 and Fig. S2 in Supplemental Material [42].

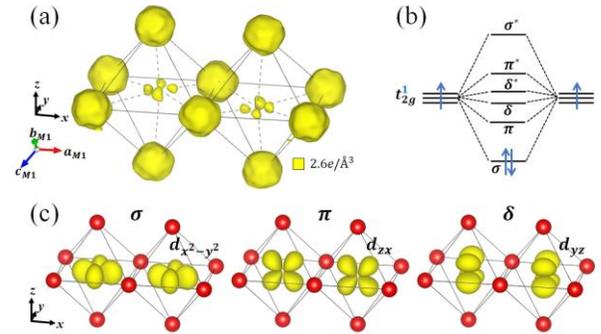

FIG. 2. (a) VED distribution of a V-V dimer and surrounding oxygen atoms obtained from the CDFS analysis at 100 K in the M1 phase. (b) Schematic of molecular orbital formation in the $t_{2g}^1$ system. (c) Calculated VED distributions of the $\sigma$, $\pi$, and $\delta$ orbitals.

To understand the VED distribution around the V site in the M1 phase [Fig. 2(a)], we consider the molecular orbital formation of $t_{2g}$ electrons. Based on the simple molecular orbital theory [52], the triplet $t_{2g}$ orbitals at the two V sites form six molecular orbitals, $\sigma$, $\pi$, $\delta$, $\delta^*$, $\pi^*$, and $\sigma^*$ orbitals in ascending order of energy [Fig. 2(b)]. Figure 2(c) shows the calculated VED distributions when each of the $\sigma$, $\pi$, and $\delta$ orbitals is occupied by two electrons. The VED obtained by the CDFS analysis has the same anisotropy as the $\sigma$-type, which is consistent with the theoretical predictions [20–23,25] and x-ray absorption spectroscopy experiments [53]. The VED around the V site is slightly higher in the direction where the dimer is formed than in other directions (Fig. S3 [42]). We succeeded in observing the molecular orbital state associated with the V-V dimerization from the VED distribution.

In the M1 phase, the V—O bond length is shortened to 1.75983(18) Å in only one direction in which the V-V dimer tilts (Fig. S4 [42]), while they are 1.92266(15) and 1.9352(3) Å in the rutile phase. Such pyramidal Jahn-Teller distortion of the $VO_6$ octahedron is also re-ported in PbVO₃ [54], stabilizing the $d_{x^2-y^2}$ orbital, as shown in Fig. S4(d) [42]. Therefore, the tilted V-V dimer arises from the molecular orbital formation and the pyramidal Jahn-Teller distortion.

Next, we focus on the VED distribution around the V site in the rutile phase, which has an anisotropy different from that of the M1 phase. The VED does not extend along the *x*-axis direction but along the *z*-axis direction where oxygen atoms exist [see (i) in Fig. 1(c)]. The VED extending toward oxygen atoms is derived from the $e_g$ orbitals, which must be empty in VO₂. To investigate the cause of the anisotropy, we carefully review the structural analysis results in the rutile phase.

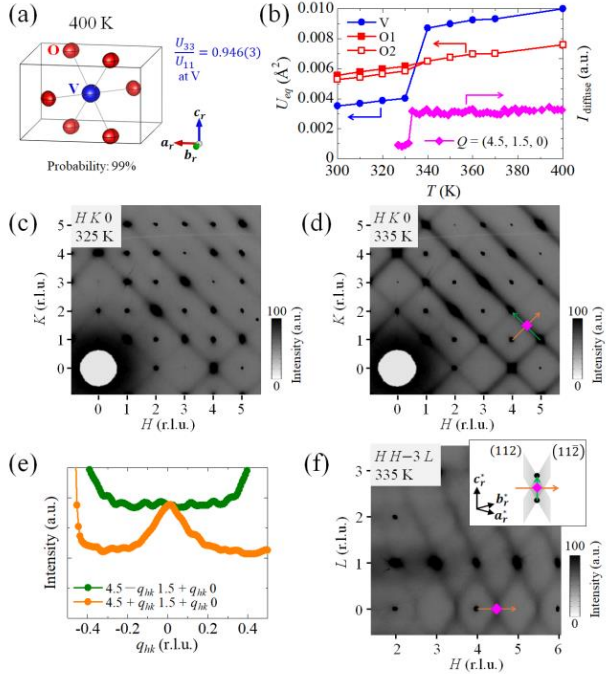

FIG. 3. (a) ADPs of VO₂ at 400 K, which are shown as ellipsoids. (b) Temperature dependence of the ADPs $U_{eq} = (U_{11} + U_{22} + U_{33})/3$ and the XDS intensity $I_{diffuse}$ at $Q = (4.5, 1.5, 0)$ in the cooling process. Below $T_{MI}$, the O site splits into the O1 and O2 sites. XRD data on the *H K* 0 plane at (c) 325 and (d) 335 K. (e) One-dimensional plots of XRD intensity around $Q = (4.5, 1.5, 0)$ along the [110] and [$\bar{1}$10] directions at 335 K, shown by orange and green dots, respectively. (f) XRD data on the *H H*–3 *L* plane at 335 K. Inset shows a schematic of the two-dimensional XDS.

Figure 3(a) shows the atomic displacement parameters (ADPs) at 400 K. There is no large anisotropy in the ADP of V [$U_{33}/U_{11} = 0.946(3)$]. Although the atomic weight of V is larger than that of O, the ADP of V is larger than that of O. This tendency is also reported by the previous structure analysis study [13]. When investigating the temperature dependence [Fig. 3(b)], the ADP of V becomes approximately halved below $T_{MI}$, whereas the ADPs of O smoothly change. The unusually large ADP of V in the rutile phase may indicate local structural fluctuations. Figures 3(c) and 3(d) show the XRD data on the *H K* 0 plane at 325 K (M1 phase) and 335 K (rutile phase), respectively. The streak like XDS is observed between Bragg peaks with $h + k + l =$ even in the rutile phase, whereas the XDS almost disappears in the M1 phase. In the rutile phase, V atoms are located at the corners and the body-center of the tetragonal unit cell, contributing only to the Bragg reflections with $h + k + l =$ even. Therefore, the XDS that only exists between Bragg peaks with $h + k + l =$ even is related to the atomic displacement of V, which is consistent with the unusually large ADP of V.

The strong XDS is observed along the Γ-M (0.5, 0.5, 0)-Γ lines in our study, whereas previous XRD measurements reported the XDS around the R (0.5, 0, 0.5) point [24,25]. Although we also observe the XDS around the R point, shown in Fig. S5(d) [42], the intensity is weaker than that around the M point. The pink squares in Fig. 3(b) show the temperature dependence of the XDS intensity at $Q = (4.5, 1.5, 0)$ corresponding to the M point. The XDS intensity suddenly drops below $T_{MI}$. The MIT temperature obtained from the XDS intensity in the cooling process is $T_{MI} = 332$ K, which is 4 K lower than the result obtained from the electrical resistivity measurements (Fig. S1 [42]). The temperature in XRD experiments appears to be underestimated by about 4 K.

The anisotropy of the XDS is investigated by one-dimensional plots along the [110] and [$\bar{1}$10] axes around $Q = (4.5, 1.5, 0)$ [a pink square in Fig. 3(d)], as shown in Fig. 3(e). Intensities are roughly constant along the [$\bar{1}$10] axis, whereas the [110] profile is characterized by a broad peak. The correlation length $\xi \sim 22$ Å, i.e., ~3.4 unit cells, along the [110] direction is estimated by the peak width. The XRD data on the *H H*–3 *L* plane at 335 K, shown in Fig. 3(f), indicates that the XDS is distributed to form planes [inset of Fig. 3(f)] but not one-dimensional streaks. In other words, there are one-dimensional correlations in the V displacement perpendicular to the XDS sheets. Furthermore, there is no peak structure at the $l = 0.5$ positions along the [112] and [$\bar{1}\bar{1}$2] directions, shown in Fig. S6(e) [42], indicating that no short-range V-V dimerization order exists in the rutile phase. If local dimers are present in the rutile phase, the effective magnetic moment in the Curie-Weiss susceptibility should be smaller than the anticipated value for $S =$ 1/2, $\mu_V = 1.73\mu_B$, owing to the formation of local singlet pairs. In fact, the effective magnetic moment $\mu_V$ is estimated to be 1.58~2.30$\mu_B$ by magnetization measurements [7], ruling out the formation of local singlet V pairs.

To understand the short-range ordered structure, the XDS pattern is simulated by V displacements. Here, the V

displacements propagating along the [111] direction are considered to reproduce the XDS corresponding to the $h + k + l = 2n$ ($n$ is an integer) plane. Figure 4(a) shows the rutile-type $VO_2$ structure showing only V atoms. Orange and green circles correspond to V1 and V2 at the (0, 0, 0) and (1/2, 1/2, 1/2) positions, respectively. Since the XDS streaks are absent at $h + k + l = 2n + 1$ on the $H$ $K$ 0 [Fig. 4(b)] and $H$ $H$ $L$ [Fig. 4(c)] planes, the V1 and V2 atoms have the same magnitude of displacements in the same direction. Furthermore, there are no XDS streaks passing through the origin, as shown in Figs. 4(b) and 4(c), indicating that the V displacements are longitudinal shifts, not transverse shifts, as shown in Fig. 4(a). Here, the calculated system size is a $32 \times 32 \times 32$ rutile cell, and the displacement magnitude of V is $\delta_r = 0.5$ Å. Random phases are assigned to the displacement directions ([111] or [$\bar{1}\bar{1}\bar{1}$]) of the one-dimensional V chain along the [111] direction, where there is no correlation between the chains.

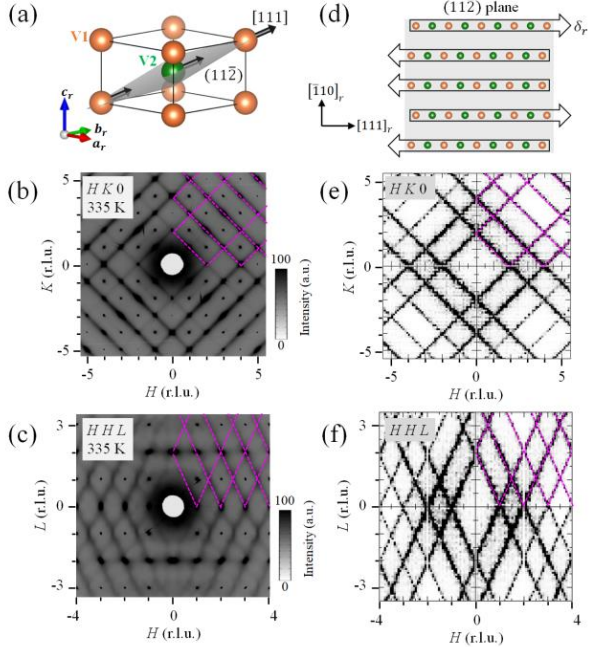

FIG. 4. (a) Schematic of the $VO_2$ structure showing only V atoms in the rutile phase. Orange and green circles correspond to V1 and V2 at the (0, 0, 0) and (1/2, 1/2, 1/2) positions, respectively. Black arrows indicate the displacements in the [111] direction. XRD data on the (b) $H$ $K$ 0 and (c) $H$ $H$ $L$ planes at 335 K in the rutile phase. (d) A part of the calculated V displacements. The calculated XDS patterns on the (e) $H$ $K$ 0 and (f) $H$ $H$ $L$ planes. Pink dotted lines indicate the XDS streaks corresponding to the experimental results in (b) and (c).

Figure 4(d) shows a part of the calculated V displacement on the $(11\bar{2})$ plane. The XDS sheets corresponding to $h + k + l = 2n$ are reproduced from the calculated structure. Figures 4(e) and 4(f) show the calculated XDS patterns on the $H$ $K$ 0 and $H$ $H$ $L$ planes considering the four domains of the tetragonal lattice (Fig. S7 [42]), which agree well with the experimental results [Figs. 4(b) and 4(c)]. The XDS patterns calculated by the longitudinal shift model are consistent with the previous XRD data [26]. The obtained short-range ordered structure does not correspond to the structural fluctuations derived from the M1 phase. In fact, the XDS sheets in the rutile phase are independent of the positions of newly appearing peaks in the M1 phase (Fig. S5 [42]). This result indicates that the phase transition in $VO_2$ is not a simple Peierls transition with softening of phonons.

The direction in which the VED extends around the V site in the rutile phase [Fig. 1(c)] corresponds to the displacement direction of V to reproduce the XDS [Fig. 4(a)]. Therefore, the VED extending in the oxygen direction would not reflect the 3$d$ orbital state, but rather the position of the core electrons of V derived from the short-range order (Fig. S8 [42]).

The short-range ordered structure is not a perfect rutile structure but a locally distorted metastable structure, which may be related to an intermediate phase suggested by the previous theoretical calculations [55,56]. The metastable intermediate phases often appear in systems exhibiting a strong first-order transition [57], which is known as an Ostwald's step rule [58]. The observed metastable structure may be related to the photoexcited transient disordering state induced [30–40] and anharmonic phonons [25] in this system.

In summary, the $t_{2g}$ molecular orbital state is observed from the VED distribution in the M1 phase of $VO_2$. The observed XDS suggests the short-range ordered structure derived from the V displacement, which is different from the structural fluctuations of the M1 phase. This metastable short-range structure may be related to peculiar properties such as the ultrafast transition and dynamics in this system.

*Acknowledgments.* We thank H. Sawa and K. Sugimoto for supporting XRD experiments, and Y. Ishii and Z. Hiroi for fruitful discussions. This work was supported by Grant-in-Aid for Scientific Research (Grants No. No. JP19H05791, No. JP22K14010, No. JP22H01166, and No. 23K13059) from JSPS. The synchrotron radiation experiments were performed at SPring-8 with the approval of the Japan Synchrotron Radiation Research Institute (JASRI) (Proposal No. 2023A1687, 2023A1882, and 2023B1603).

*kitou@edu.k.u-tokyo.ac.jp

# Supplemental Material of

# Molecular orbital formation and metastable short-range ordered structure in VO$_2$

Shunsuke Kitou[1], Akitoshi Nakano[2], Masato Imaizumi[2], Yuiga Nakamura[3],

Ichiro Terasaki[2], Taka-hisa Arima[1,4]

[1]*Department of Advanced Materials Science, The University of Tokyo, Kashiwa 277-8561, Japan.*

[2]*Department of Physics, Nagoya University, Nagoya 464-8602, Japan.*

[3]*Japan Synchrotron Radiation Research Institute (JASRI), SPring-8; Hyogo 679-5198, Japan.*

[4]*RIKEN Center for Emergent Matter Science, Wako 351-0198, Japan.*

Table S1. Structural parameters of VO$_2$ at 400 K. The space group is $P4_2/mnm$ (No. 136) and $a = 4.5571(6)$Å, $c = 2.8565(2)$Å. Note that $U_{11} = U_{22}$ and $U_{13} = U_{23} = 0$ for 2a and 4g sites.

| Atom | Wyckoff position | Site symmetry | $x$ | $y$ | $z$ |
|------|------------------|---------------|-----|-----|-----|
| V | 2a | m.mm | 0 | 0 | 0 |
| O | 4g | m.2m | 0.30028(2) | 0.69972(2) | 0 |

| Atom | $U_{11}$ (Å$^2$) | $U_{33}$ (Å$^2$) | $U_{12}$ (Å$^2$) |
|------|------------------|------------------|------------------|
| V | 0.01018(2) | 0.00963(3) | -0.001080(10) |
| O | 0.00784(3) | 0.00713(4) | 0.00214(3) |

Table S2. Summary of crystallographic data of VO$_2$ at 400 K.

| | |
|---|---|
| Wavelength (Å) | 0.30946 Å |
| Crystal dimension ($\mu m^3$) | 45×40×30 |
| Space group | $P4_2/mnm$ |
| $a$ (Å) | 4.5571(6) |
| $c$ (Å) | 2.8565(2) |
| Z | 2 |
| $F(000)$ | 78 |
| $(\sin\theta/\lambda)_{max}$ (Å$^{-1}$) | 1.79 |
| $N_{total}$ | 8509 |
| $N_{unique}$ ($\sin\theta/\lambda > 0.6$Å$^{-1}$ / all) | 763 / 802 |
| Average redundancy | 10.61 |
| Completeness (%) | 97.33 |
| $N_{parameters}$ | 8 |
| $R_1$ ($\sin\theta/\lambda > 0.6$Å$^{-1}$ / all) | 2.50% / 3.82% |
| $wR_2$ ($\sin\theta/\lambda > 0.6$Å$^{-1}$ / all) | 3.01% / 3.46% |
| GOF ($\sin\theta/\lambda > 0.6$Å$^{-1}$ / all) | 1.32 / 1.54 |

Table S3. Structural parameters of VO$_2$ at 100 K. The space group is $P2_1/c$ (No. 14) and $a = 5.7394(2)$Å, $b = 4.5255(2)$Å, $c = 5.3801(2)$Å, $\beta = 122.566(9)°$.

| Atom | Wyckoff position | Site symmetry | $x$ | $y$ | $z$ |
|------|------|------|------|------|------|
| V | 4$e$ | 1 | 0.761020(6) | 0.978198(7) | 0.472918(6) |
| O(1) | 4$e$ | 1 | 0.59920(3) | 0.70245(3) | 0.20103(3) |
| O(2) | 4$e$ | 1 | 0.89350(3) | 0.21257(3) | 0.29085(3) |

| Atom | $U_{11}$ (Å$^2$) | $U_{22}$ (Å$^2$) | $U_{33}$ (Å$^2$) | $U_{12}$ (Å$^2$) | $U_{13}$ (Å$^2$) | $U_{23}$ (Å$^2$) |
|------|------|------|------|------|------|------|
| V | 0.002211(9) | 0.002103(10) | 0.002122(8) | 0.000056(4) | 0.001229(6) | -0.000010(4) |
| O(1) | 0.00352(3) | 0.00357(3) | 0.00361(3) | -0.00063(2) | 0.00191(2) | -0.00080(2) |
| O(2) | 0.00317(3) | 0.00358(3) | 0.00354(3) | 0.00035(2) | 0.00187(2) | 0.00088(2) |

Table S4. Summary of crystallographic data of VO$_2$ at 100 K.

| | |
|------|------|
| Wavelength (Å) | 0.30946 Å |
| Crystal dimension ($\mu m^3$) | 46×39×29 |
| Space group | $P2_1/c$ |
| $a$ (Å) | 5.7394(2) |
| $b$ (Å) | 4.5255(2) |
| $c$ (Å) | 5.3801(2) |
| $\beta$ (°) | 122.566(9) |
| Z | 4 |
| $F(000)$ | 156 |
| $(\sin\theta/\lambda)_{max}$ (Å$^{-1}$) | 1.79 |
| $N_{total}$ | 29107 |
| $N_{unique}$ ($\sin\theta/\lambda$>0.6Å$^{-1}$ / all) | 5244 / 5456 |
| Average redundancy | 5.335 |
| Completeness (%) | 95.35 |
| $N_{parameters}$ | 28 |
| $R_1$ ($\sin\theta/\lambda$>0.6Å$^{-1}$ / all) | 1.75% / 1.92% |
| $wR_2$ ($\sin\theta/\lambda$>0.6Å$^{-1}$ / all) | 2.65% / 2.80% |
| GOF ($\sin\theta/\lambda$>0.6Å$^{-1}$ / all) | 1.49 / 1.60 |

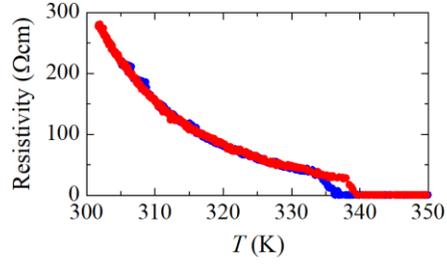

FIG. S1. Temperature dependence of the electric resistivity of VO$_2$. The phase transition temperatures are 340 and 336 K during the heating and cooling processes, which are consistent with previous resistivity measurements [1-3].

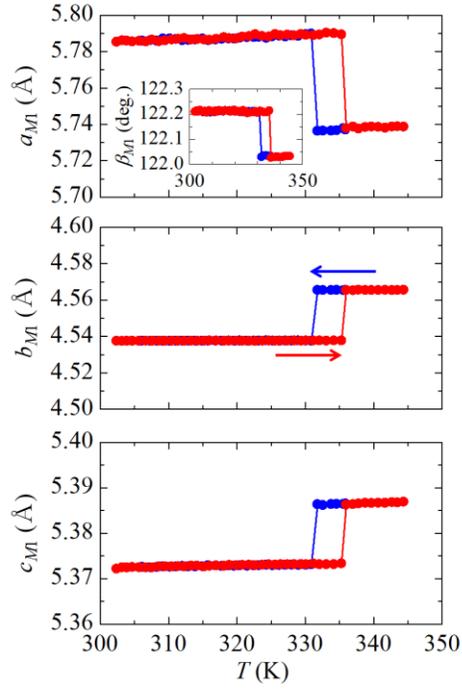

FIG. S2. Temperature dependence of the lattice parameters near the metal-insulator transition of VO$_2$. Blue and red dots indicate the cooling and heating process, respectively.

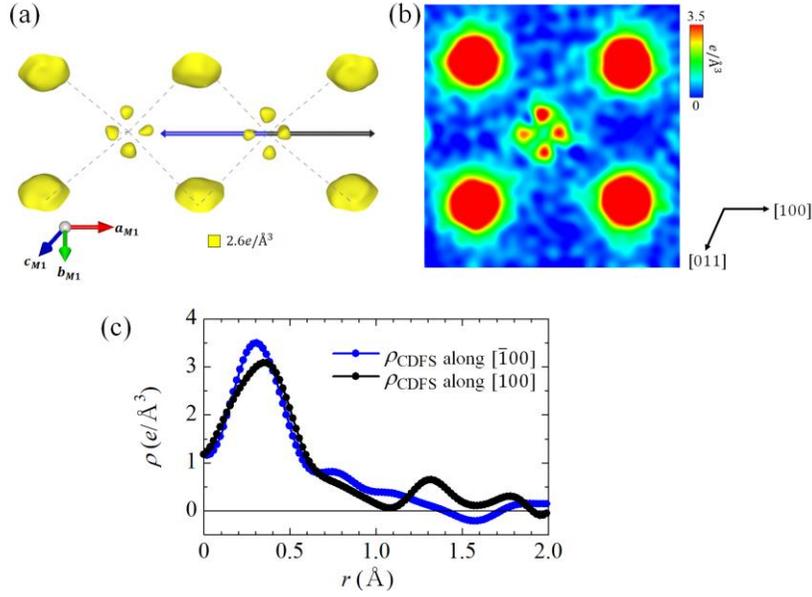

FIG. S3. (a) VED distributions obtained from the CDFS analysis at 100 K. The V-V dimer is formed along the blue vector direction. (b) A two-dimensional color plot of the VED of the $(01\bar{1})$ plane on V, which corresponds to the right side one of (a). (c) Blue and Black dots indicate the one-dimensional plots of the VED as functions of the distance $r$ from the V nucleus along the $[\bar{1}00]$ and $[100]$ directions, respectively, which correspond to the blue and black vectors in (a).

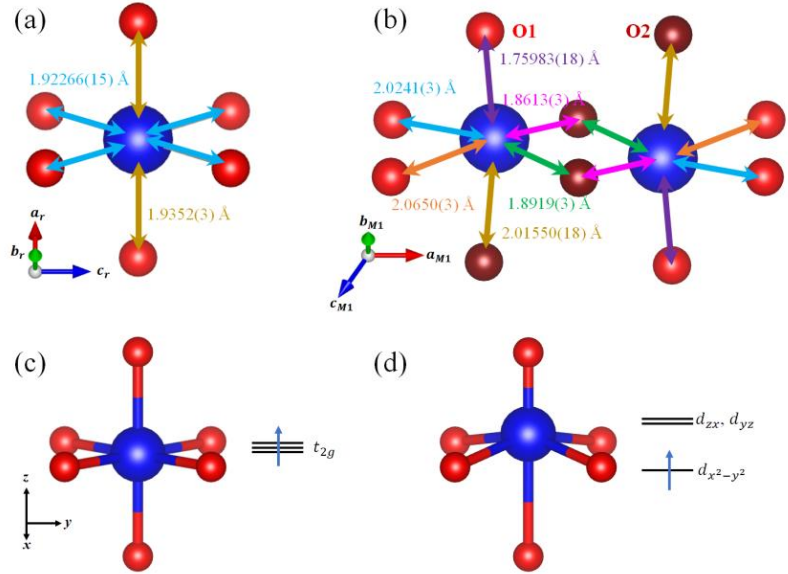

FIG. S4. The V—O bond lengths in $VO_6$ octahedra at (a) 400 K in the rutile phase and (b) 100 K in the M1 phase. Arrows of same color represent the same bond length. Schematic of the $VO_6$ octahedron and the $t_{2g}^1$ orbital state (c) without distortion and (d) with pyramidal distortion.

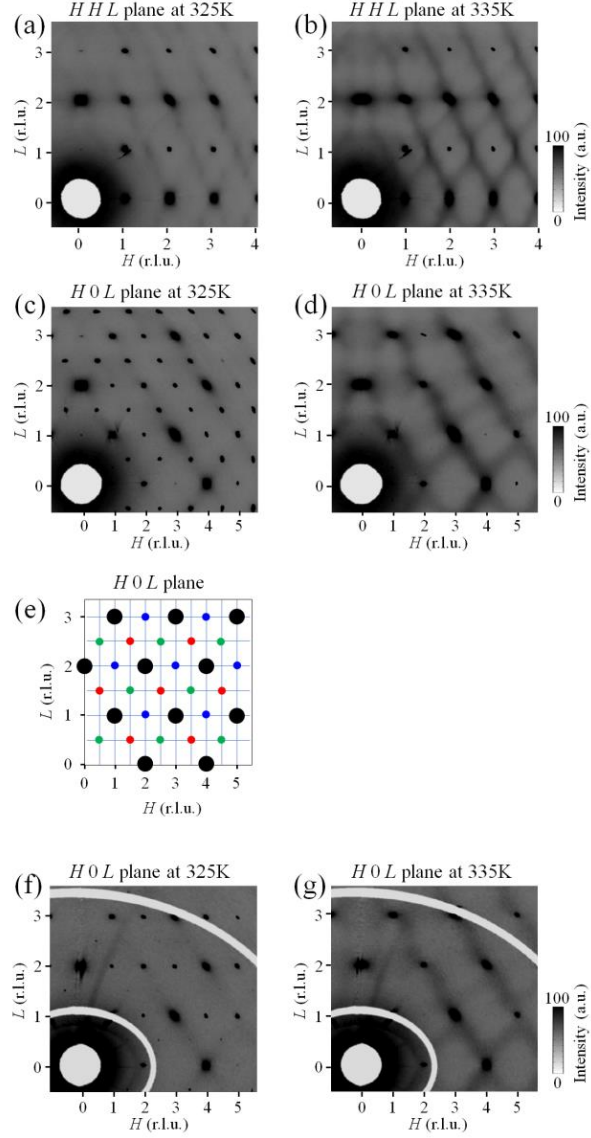

FIG. S5. XRD data for the larger crystal mentioned in the main manuscript in the $H\,H\,L$ planes at (a) 325 and (b) 335 K, and the $H\,0\,L$ planes at (c) 325 and (d) 335 K, respectively. (e) Schematic of the XRD pattern with monoclinic domains in the $H\,0\,L$ plane. Black circles indicate Bragg peaks in the rutile phase. Blue, green, red circles indicate newly appearing peaks in the M1 phase, which correspond to the domains of $\boldsymbol{a}_{M1} = 2\boldsymbol{c}_R$, $\boldsymbol{b}_{M1} = \boldsymbol{b}_R$, $\boldsymbol{c}_{M1} = -\boldsymbol{a}_R - \boldsymbol{c}_R$, $\boldsymbol{a}'_{M1} = 2\boldsymbol{c}_R$, $\boldsymbol{b}'_{M1} = \boldsymbol{a}_R$, $\boldsymbol{c}'_{M1} = \boldsymbol{b}_R - \boldsymbol{c}_R$, and $\boldsymbol{a}''_{M1} = 2\boldsymbol{c}_R$, $\boldsymbol{b}''_{M1} = -\boldsymbol{a}_R$, $\boldsymbol{c}''_{M1} = -\boldsymbol{b}_R - \boldsymbol{c}_R$, respectively. XRD data for another crystal measuring $210 \times 50 \times 30\ \mu m^3$ with nearly a single domain in the $H\,0\,L$ planes at (f) 325 K and (g) 335 K, respectively. The newly appearing peak positions in the M1 phase are independent of the XDS streaks in the rutile phase.

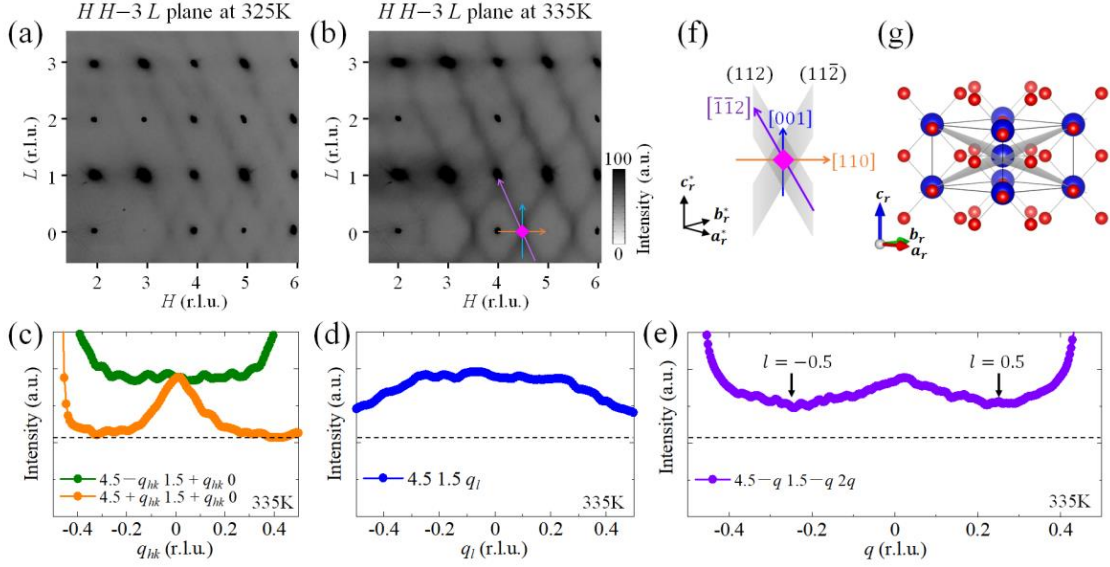

FIG. S6. XRD data for the larger crystal mentioned in the main manuscript in the $H\,H$–3 $L$ plane at (a) 325 and (b) 335 K. One-dimensional plots of XRD intensity along the (c) [110], [$\bar{1}$10], (d) [001], and (e) [$\bar{1}\bar{1}$2] directions at 335 K. Black dotted lines in (c)-(e) correspond to the background intensity. (f) Schematic of the two-dimensional diffuse scattering planes of (112) and (11$\bar{2}$). (g) The scattering planes corresponding to (112) and (11$\bar{2}$) of the $VO_2$ structure in the rutile phase.

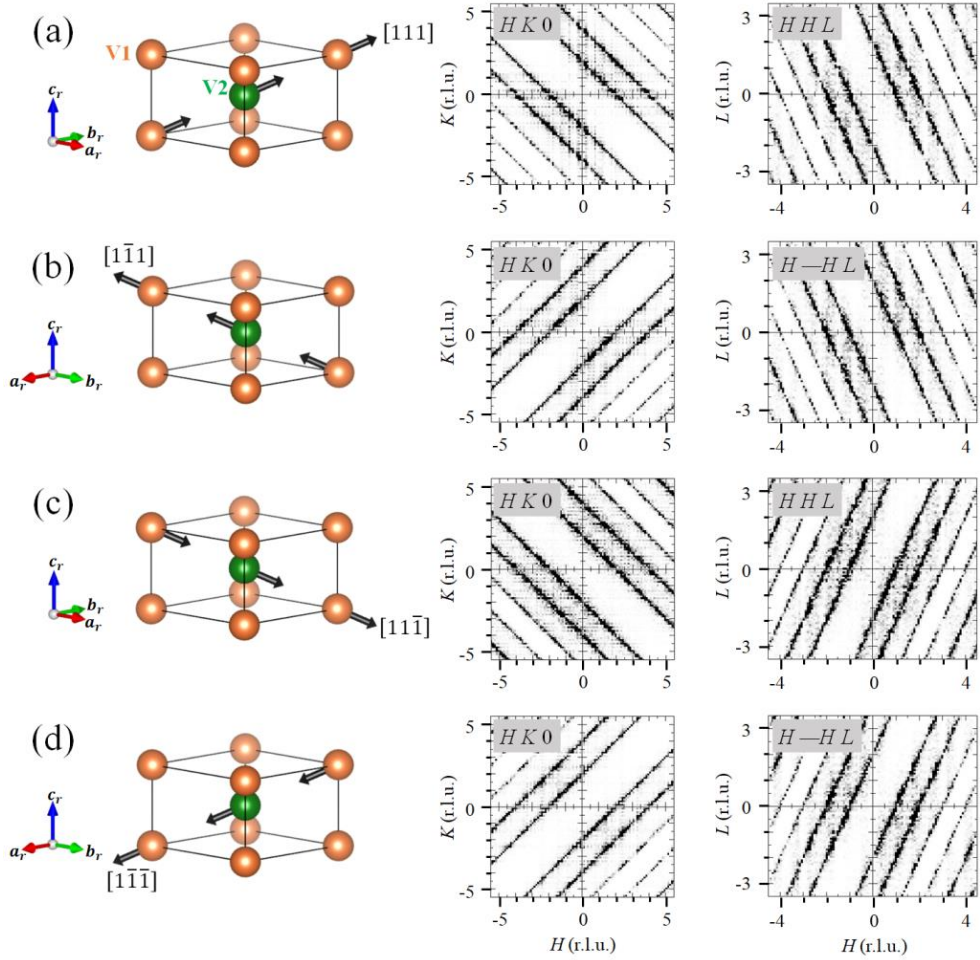

FIG. S7. Schematic of the VO$_2$ structures and the diffuse scattering patterns on the $H\,K\,0$ and $H\,H$ (or $-H$) $L$ planes corresponding to the short-range ordering along the (a) [111], (b) [1$\bar{1}$1], (c) [11$\bar{1}$], and (d) [1$\bar{1}\bar{1}$] directions.

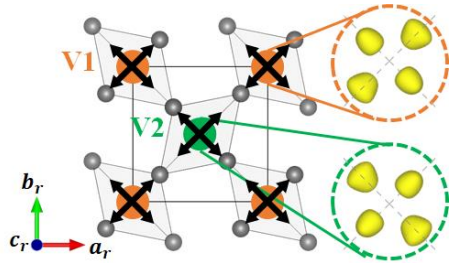

FIG. S8. VED distributions around the V1 and V2 sites at 400 K in the rutile phase. Black arrows indicate the V displacement corresponding to the short-range order.